\documentclass[aps,preprint,amsmath]{revtex4}

\usepackage{graphicx}
\usepackage{pstricks}

\def\al{\alpha}
\def\be{\beta}

\def\ep{\epsilon}
\def\ve{\varepsilon}
\def\ze{\zeta}
\def\et{\eta}
\def\th{\theta}

\def\ka{\kappa}
\def\la{\lambda}

\def\rh{\rho}

\def\si{\sigma}
\def\vs{\varsigma}

\def\ph{\phi}

\def\om{\omega}

\def\De{\Delta}

\def\La{\Lambda}

\def\Om{\Omega}

\def\cL{{\cal L}}

\def\fr#1#2{{{#1} \over {#2}}}
\def\half{{\textstyle{1\over 2}}}

\def\frac#1#2{{\textstyle{{#1}\over {#2}}}}

\def\lsim{\mathrel{\rlap{\lower4pt\hbox{\hskip1pt$\sim$}}
    \raise1pt\hbox{$<$}}}
\def\gsim{\mathrel{\rlap{\lower4pt\hbox{\hskip1pt$\sim$}}
    \raise1pt\hbox{$>$}}}
\def\sqr#1#2{{\vcenter{\vbox{\hrule height.#2pt
         \hbox{\vrule width.#2pt height#1pt \kern#1pt
         \vrule width.#2pt}
         \hrule height.#2pt}}}}

\def\prt{\partial}

\def\etal{{\it et al.}}

\newcommand{\beq}{\begin{equation}}
\newcommand{\eeq}{\end{equation}}
\newcommand{\bea}{\begin{eqnarray}}
\newcommand{\eea}{\end{eqnarray}}
\newcommand{\bit}{\begin{itemize}}
\newcommand{\eit}{\end{itemize}}
\newcommand{\rf}[1]{(\ref{#1})}

\def\cv{\v{C}erenkov}

\def\mn{{\mu\nu}}

\def\kl{{\ka\la}}

\def\rs{{\rh\si}}

\newcounter{tc1}\newcounter{tc2}
\newcounter{tr1}\newcounter{tr2}
\newlength{\h}
\def\newtableau#1#2{\psset{unit=12pt,linewidth=0.5pt}%
  \setlength{\h}{#2\psunit}\setlength{\h}{0.5\h}\addtolength{\h}{-0.3\psunit}
  \begin{pspicture}[shift=-\h](#1,#2)\small%
    \setcounter{tc1}{0}\setcounter{tc2}{1}%
    \setcounter{tr1}{#2}\setcounter{tr2}{#2}\addtocounter{tr1}{-1}%
    \psline(0,0)(0,#2)(#1,#2)}
\def\endtableau{\end{pspicture}}
\def\newbox#1{%
  \psline(\value{tc1},\value{tr1})(\value{tc2},\value{tr1})(\value{tc2},\value{tr2})%
  \rput(\value{tc1},\value{tr1}){\rput(0.5,0.5){#1}}
  \addtocounter{tc1}{1}\addtocounter{tc2}{1}}
\def\longbox#1{%
  \addtocounter{tc2}{1}%
  \psline(\value{tc1},\value{tr1})(\value{tc2},\value{tr1})(\value{tc2},\value{tr2})%
  \rput(\value{tc1},\value{tr1}){\rput(1.0,0.5){#1}}
  \addtocounter{tc1}{2}\addtocounter{tc2}{1}}
\def\newrow{%
  \addtocounter{tr1}{-1}\addtocounter{tr2}{-1}%
  \setcounter{tc1}{0}\setcounter{tc2}{1}}

\def\K{\mathcal K}

\def\Kd{{\K}^{(d)}{}}
\def\KHat{\widehat{\K}^{(d)}{}}

\def\b{q}
\def\bHat{\widehat{\b}}
\def\bd#1{{\b}^{(#1)}}

\def\c{s}
\def\cHat{\widehat{\c}}
\def\cd#1{{\c}^{(#1)}}

\def\ct{\overline\c}
\def\ctHat{\skew{3}\widehat{\ct}}

\def\d{k}
\def\dHat{\widehat{\d}}
\def\dd#1{{\d}^{(#1)}}

\def\dc{\circ}

\def\qrt{\tfrac14}

\def\mbf#1{\mbox{\boldmath$#1$}}
\def\pvec{\mbf p}

\def\nhat{\mbf{\hat n}}

\def\syjm#1#2{\phantom{}_{#1}Y_{#2}}

\def\kjm#1#2#3{k^{(#1)}_{(#2)#3}}
\def\kI{\kjm{d}{I}{jm}}
\def\kE{\kjm{d}{E}{jm}}
\def\kB{\kjm{d}{B}{jm}}
\def\kV{\kjm{d}{V}{jm}}

\def\kIdjm#1#2{\kjm{#1}{I}{#2}}
\def\kEdjm#1#2{\kjm{#1}{E}{#2}}
\def\kBdjm#1#2{\kjm{#1}{B}{#2}}
\def\kVdjm#1#2{\kjm{#1}{V}{#2}}

\def\ring#1{{\mathaccent'27 #1}}
\def\kring#1#2{\ring k^{(#1)}_{(#2)}}
\def\kIring{\kring{d}{I}}
\def\kVring{\kring{d}{V}}
\def\kIdring#1{\kring{#1}{I}}
\def\kVdring#1{\kring{#1}{V}}

\begin{document}

\title{Testing local Lorentz invariance with gravitational waves
}

\author{V.\ Alan Kosteleck\'y$^1$ and Matthew Mewes$^2$}

\affiliation{$^1$Physics Department, Indiana University,
Bloomington, Indiana 47405, USA\\
$^2$Physics Department, California Polytechnic State University, 
San Luis Obispo, California 93407, USA}

\date{IUHET 611, February 2016}

\begin{abstract}

The effects of local Lorentz violation on 
dispersion and birefringence of gravitational waves are investigated.
The covariant dispersion relation for gravitational waves
involving gauge-invariant Lorentz-violating operators 
of arbitrary mass dimension is constructed.
The chirp signal from the gravitational-wave event GW150914 
is used to place numerous first constraints 
on gravitational Lorentz violation.

\end{abstract}

\maketitle

The recent discovery of gravitational waves
\cite{ligo},
a century after their prediction by Einstein
\cite{einstein},
opens the door to a new class of experimental tests 
of General Relativity (GR).
While GR is an impressively successful 
classical field theory of gravity,
incorporating a consistent description of quantum effects 
is widely expected to involve changes to its underlying principles.
An essential foundation of GR 
is the Einstein equivalence principle,
which combines the requirements of local Lorentz invariance
with local position invariance
and the weak equivalence principle.
In this work,
we demonstrate that the observation of gravitational waves
from coalescing black holes at cosmological distances
presents an opportunity for clean tests of local Lorentz invariance
in the pure-gravity sector.
We use the chirp data from the gravitational-wave event GW150914
to place first constraints on certain types of local Lorentz violation
involving the gravitational field.

Experimental studies of local Lorentz invariance,
which includes symmetry under local rotations and boosts,
have enjoyed a resurgence in popularity  
in recent decades
\cite{tables,will},
triggered by the demonstration
that minuscule Lorentz violation
could naturally emerge in quantum-gravity theories
such as strings
\cite{ksp}.
A general and model-independent approach
to describing the effects of Lorentz violation in quantum gravity
is provided by effective field theory
\cite{akgrav}.
We are interested here in the corresponding action for pure gravity,
which is formed as the sum 
of the usual Einstein-Hilbert action with cosmological constant 
together with all possible terms 
involving operators formed from gravitational fields.
This theory is a piece of the general effective field theory
for gravity and matter,
the gravitational Standard-Model Extension (SME).
A Lorentz-violating term in the action 
is an observer-independent quantity 
containing a Lorentz-violating operator 
contracted with a coefficient governing the size of its effects.
Using natural units,
each operator can be assigned a mass dimension $d$,
and the associated coefficient then has mass dimension $4-d$.
Under the plausible assumption that Lorentz violation
is suppressed by powers of the Planck mass,
which is the natural mass scale 
associated with the Newton gravitational coupling,
operators of higher $d$ can be expected to induce smaller effects. 

Laboratory
\cite{2007Battat,2007MullerInterf,2009Chung,%
2010Bennett,2012Iorio,2013Bailey,2014Shao,he15}
and analytical 
\cite{bk,se09,al10,kt11,bt11,jt12,yb15,je15}
investigations using this pure-gravity effective field theory
have largely concentrated on minimal Lorentz-violating operators,
which have dimension $d=4$.
Theoretical aspects of nonminimal operators
with dimensions $d=5,6$ have been studied 
\cite{bkx},
and constraints on some nonrelativistic combinations of operators with $d=6$
have been obtained via laboratory tests of short-range gravity
\cite{lo15,hust15,hustiu}.
The tighest constraints to date
on local Lorentz violation in the gravity sector
have been deduced from the absence of gravitational \cv\ radiation
by cosmic rays
\cite{ca80,mo01,el05,ki12,la12,kt15,ki15}
with a large class of effects for $d=4$, $6$, and $8$
now being excluded at sharp levels
\cite{kt15}.
Reviews can be found, 
for example,
in Refs.\ \cite{tables,will,jt,li13,rb}. 
Here,
we construct the general quadratic Lagrange density 
for gravitational waves 
in the presence of Lorentz-violating operators of arbitrary $d$,
and we extract the covariant dispersion relation 
involving gauge-invariant effects.
We show that observations of gravitational waves
provide sensitivities to nonminimal Lorentz violation 
independent of matter-sector effects,
and we use the gravitational-wave event GW150914 
to place numerous first limits on gravity-sector operators with $d\geq 5$. 
 
The effective field theory for gravitational Lorentz violation
\cite{akgrav}
can be linearized 
in a flat-spacetime background with Minkowski metric,
$g_\mn = \et_\mn+ h_\mn$.
Our first goal is to construct the general Lagrange density
quadratic in the dimensionless metric perturbation $h_\mn$,
allowing for both Lorentz-invariant and Lorentz-violating terms.
A generic term of this type takes the form
\beq
\cL_{\Kd} = \qrt h_\mn \KHat^{\mn\rs} h_\rs ,
\label{L1}
\eeq
where
\bea
\KHat^{\mn\rs}
&=&
\Kd^{\mn\rs\ve_1\ve_2\ldots\ve_{d-2}}
\prt_{\ve_1} \prt_{\ve_2}\ldots \prt_{\ve_{d-2}}
\nonumber\\
&\equiv &
\Kd^{\mn\rs{\dc^{d-2}}}
\label{KHat}
\eea
is an operator of mass dimension $d\geq 2$
and the coefficients $\Kd^{\mn\rs\ve_1\ve_2\ldots\ve_{d-2}}$
have mass dimension $4-d$ and are assumed constant and small.
We have introduced here a convenient notation 
by which indices contracted into a derivative are denoted
with a circle index $\dc$,
and $n$-fold contractions are denoted as $\dc^n$.
The Lorentz-invariant pieces of the expression \rf{KHat}
consist of complete traces of the coefficients 
$\Kd^{\mn\rs\ve_1\ve_2\ldots\ve_{d-2}}$.
Varying the action 
reveals that only operators satisfying the condition
$\KHat^{(\mu\nu)(\rh\si)} \pm \KHat^{(\rh\si)(\mu\nu)} \neq 0$
can contribute to the equations of motion,
where the upper sign is for even $d$ and lower one for odd $d$.

\begin{table*}
\def\vsp{&&&\\[-8pt]}
\tabcolsep4pt
\begin{tabular}{c|l|c|c|c}
Tableau & \multicolumn{1}{c|}{Operator $\KHat^{\mn\rs}$} 
& CPT 
& $d$ 
& Number 
\\
\hline\hline
\vsp
$\newtableau{4}{3}
\newbox{$\mu$}\newbox{$\nu$}\longbox{$\cdots$}
\newrow\newbox{$\rh$}\newbox{$\si$}
\newrow\newbox{$\dc$}\newbox{$\dc$}
\endtableau$
& $\cd{d}{}^{\mu\rh\dc\nu\si\dc\dc^{d-4}}$
& even
& even, $\geq 4$ & $(d-3)(d-2)(d+1)$ 
\\
\vsp
\hline
\vsp
$\newtableau{5}{3}
\newbox{$\mu$}\newbox{$\nu$}\newbox{$\si$}\longbox{$\cdots$}
\newrow\newbox{$\rh$}\newbox{$\dc$}\newbox{$\dc$}
\newrow\newbox{$\dc$}
\endtableau$
& $\bd{d}{}^{\mu\rh\dc\nu\dc\si\dc\dc^{d-5}}$
& odd 
& odd, $\geq 5$ & $\tfrac52(d-4)(d-1)(d+1)$ 
\\
\vsp
\hline
\vsp
$\newtableau{6}{2}
\newbox{$\mu$}\newbox{$\nu$}\newbox{$\rh$}\newbox{$\si$}\longbox{$\cdots$}
\newrow\newbox{$\dc$}\newbox{$\dc$}\newbox{$\dc$}\newbox{$\dc$}
\endtableau$
& $\dd{d}{}^{\mu\dc\nu\dc\rh\dc\si\dc\dc^{d-6}}$
& even
& even, $\geq 6$ & $\tfrac52(d-5)d(d+1)$
\end{tabular}
\caption{\label{table1}
Gauge-invariant operators in the quadratic gravitational action.}
\end{table*}

To construct explicitly the operators 
$\KHat^{\mn\rs}$,
we perform a decomposition into irreducible pieces
and examine the properties of each.
This reveals that 14 independent classes of operators
can control the behavior of gravitational waves.
However,
many violate the usual gauge symmetry of GR
under the transformaton 
$h_\mn \to h_\mn + \prt_\mu \xi_\nu + \prt_\nu \xi_\mu$.
Performing this transformation on the term \rf{KHat}
shows that the condition for gauge invariance is 
$\big(\KHat^{(\mu\nu)(\rh\si)} 
\pm \KHat^{(\rh\si)(\mu\nu)}\big) \prt_{\nu} = 0$.
Only three of the 14 classes of irreducible operators
obey this condition,
and they are therefore of particular interest. 
Their existence can be understood as following
from spontaneous breaking of the diffeomorphism and Lorentz invariance 
\cite{bk05},
which hides symmetry rather than explicitly violating it
and is therefore automatically compatible with the Bianchi identities
\cite{akgrav}.
The gauge invariance maintains 
the standard counting of degrees of freedom in $h_\mn$,
insuring that the three classes of operators
induce perturbative modifications 
to the two usual propagating modes in a gravitational wave.
Note that in principle
the higher derivatives occuring in the term \rf{KHat}
introduce additional modes,
but these are nonperturbative in Lorentz violation 
and occur only at high energies 
outside the domain of validity of the effective field theory. 

The three classes of gauge-invariant operators 
are determined by their symmetries,
which are given by the Young tableaux in Table \ref{table1}.
It is convenient to denote them by the three specific symbols
shown in the second column of the table,
instead of the generic form $\KHat^{\mn\rs}$.
The CPT handedness of the corresponding terms 
in the quadratic Lagrange density 
is given in the third column.
The operators exist only in the dimensions listed 
in the fourth column of the table.
The number of independent components of each
is displayed in the fifth column of the table.
In what follows,
it is also useful to define quantities
that are the sums over $d$ of each of these sets of operators,
\bea
\cHat{}^{\mu\rh\nu\si} &=& 
\sum_d \cd{d}{}^{\mu\rh\dc\nu\si\dc^{d-3}} ,
\quad
\bHat{}^{\mu\rh\nu\si} = 
\sum_d \bd{d}{}^{\mu\rh\dc\nu\dc\si\dc^{d-4}},
\notag \\
&&
\hskip 30pt
\dHat{}^{\mu\nu\rh\si} = 
\sum_d \dd{d}{}^{\mu\dc\nu\dc\rh\dc\si\dc^{d-5}} .
\label{coeffs}
\eea
The operator $\bHat{}^{\mu\rh\nu\si}$
is antisymmetric in the first pair of indices
and symmetric in the second,
while the operator $\cHat{}^{\mu\rh\nu\si}$
is antisymmetric in both the first and second pairs of indices,
and $\dHat{}^{\mu\nu\rh\si}$ is totally symmetric.
Any contraction of these operators with a derivative vanishes.

The complete gauge-invariant quadratic Lagrange density, 
including all Lorentz-violating and Lorentz-invariant terms
of arbitrary mass dimension $d$,
then takes the form
\bea
\cL &=& \cL_0 + 
\qrt h_\mn 
(\cHat{}^{\mu\rh\nu\si} 
+ \bHat{}^{\mu\rh\nu\si} 
+ \dHat{}^{\mu\nu\rh\si})
h_\rs,
\nonumber\\
\cL_{0} &=& 
\qrt \ep^{\mu\rh\al\ka} \ep^{\nu\si\be\la} 
\et_\kl h_\mn \prt_\al\prt_\be h_\rs,
\label{lag}
\eea
where $\cL_0$
is the quadratic approximation to the Einstein-Hilbert action
and is a subset of the $d=4$ component of $\cHat{}^{\mu\rh\nu\si}$.
We remark in passing that the introduction of a dual operator via
$\cHat{}^{\mu\rh\nu\si}
=-\ep^{\mu\rh\al\ka}\ep^{\al\nu\si\be\la}\ctHat{}_\kl \prt_\al \prt_\be$
reveals that $\cHat{}^{\mu\rh\nu\si}$ contributes
as a momentum-dependent metric perturbation,
$\et_\kl \to \et_\kl-\ctHat{}_\kl$.
The effects of $\ctHat{}_\kl$ 
on gravitational \cv\ radiation is the subject
of Ref.\ \cite{kt15}. 
Also,
a treatment of terms linear in $h_\mn$
is provided elsewhere,
along with a discussion of the other 11 classes 
of gauge-violating operators,
which can describe additional gravitational modes
beyond the usual helicity-two states of GR
\cite{km16}.

Following methods developed for the study of Lorentz invariance
in the photon sector of the SME
\cite{km09},
the covariant dispersion relation for propagation 
of a gravitational wave of 4-momentum $p^\mu = (\om, \mbf p)$
can be derived from the Lagrange density \rf{lag}.
Some algebra reveals that the leading-order dispersion relation
takes the form
\beq
\om = \Big(
1-\vs^0 \pm \sqrt{(\vs^1)^2+(\vs^2)^2+(\vs^3)^2}\,\Big)|\pvec|, 
\label{disprel}
\eeq
with 
\bea
\vs^0 &=& \frac{1}{4\pvec^2}\Big(
- \cHat{}^{\mn}{}_{\mn}
+\half \dHat{}^{\mu\nu}{}_{\mu\nu} \Big),
\notag \\
(\vs^1)^2+(\vs^2)^2 &=&
\frac{1}{8\pvec^4}\Big(
\dHat{}^{\mu\nu\rh\si}\dHat{}_{\mu\nu\rh\si}
-\dHat{}^{\mu\rh}{}_{\nu\rh}\dHat{}_{\mu\si}{}^{\nu\si}
\notag \\
&&
\hskip 30pt
+\tfrac{1}{8} \dHat{}^{\mu\nu}{}_{\mu\nu} \dHat{}^{\rh\si}{}_{\rh\si}
\Big) ,
\notag \\
(\vs^3)^2 &=& \frac{1}{16\pvec^4}\Big(
-\half \bHat{}^{\mu\rh\nu\si}\bHat{}_{\mu\rh\nu\si}
-\bHat{}^{\mu\nu\rh\si}\bHat{}_{\mu\rh\nu\si}
\notag \\
&&
\hskip 30pt
+\big(\bHat{}^{\mu\rh\nu}{}_{\rh}+\bHat{}^{\nu\rh\mu}{}_{\rh}\big)
\bHat{}_{\mu\si\nu}{}^{\si}
\Big) ,
\qquad
\eea
where now the derivative factors $\prt_\mu$
in the coefficients \rf{coeffs}
are understood to be replaced by their 4-momentum equivalent,
$\prt_\mu \to i p_\mu$.

The structure of the dispersion relation \rf{disprel}
indicates that Lorentz-violating modifications
to the propagation of gravitational waves
can be classified in terms of 
anisotropy, dispersion, and birefringence.
Anisotropy is a consequence of the breaking of rotation symmetry,
and in a specified observer frame
it is controlled by coefficients 
for Lorentz violation with spatial indices.
All three classes of gauge-invariant operators
can produce anisotropic effects.
Dispersion arises when the speed of the gravitational wave
depends on its frequency.
Since every coefficient for Lorentz violation
with $d>4$ is associated with powers of momenta
in the dispersion relation,
only coefficients with $d=4$ produce dispersion-free propagation.
These are all contained in $\cHat{}^{\mu\rh\nu\si}$.
Finally,
the separation of polarization modes evident in the dispersion relation
through the two branches of the square root
implies that birefringence of gravitational waves
can be caused only by the operators 
$\bHat{}^{\mu\rh\nu\si}$ and $\dHat{}^{\mu\rh\nu\si}$
and hence only for $d>4$.

A gravitational wave traveling along $\mbf{\hat p}$
and detected by a laboratory in the vicinity of the Earth 
appears to emanate 
from a source located in the direction of the unit radial vector 
$\mbf{\hat n}=-\mbf{\hat p}$
in spherical polar coordinates centered on the Earth.
For example,
the most likely location of the source of
the gravitational-wave event GW150914 
is a region of about $\sim 600$ square degrees
in the southern hemisphere
around declination $-70^\circ$
and right ascension $8$ hr
\cite{ligo2},
so that $\nhat$ has spherical polar angles
$\th\simeq 160^\circ$, 
$\ph \simeq 120^\circ$
in the Sun-centered celestial-equatorial frame
canonically used to report results of searches for Lorentz violation
\cite{sunframe}.
For practical applications,
it is therefore useful to perform a decomposition
of the dispersion relation \rf{disprel}
in spherical harmonics.
Since the metric perturbation $h_\mn$ has helicity-2 components,
and since the Lagrange density \rf{lag}
is quadratic in $h_\mn$,
the decomposition involves spin-weighted spherical harmonics 
\citep{sYjm}
of spin weight $s\leq 4$,
which we denote as $\syjm{s}{jm}(\mbf{\hat n})$.
A summary of the properties of these harmonics can be found
in Appendix A of Ref.\ \cite{km09}.
Note in particular that 
$\syjm{0}{jm}(\mbf{\hat n}) \equiv Y_{jm} (\mbf{\hat n})$,
the usual scalar spherical harmonics.

Investigation shows that
$\vs^0$, $\vs^3$ are rotation scalars
while $\vs^1$, $\vs^2$ are helicity-4 tensors.
The decomposition then can be written as 
\bea
\vs^0 &=& 
\sum_{djm} \om^{d-4} \, \syjm{}{jm}(\mbf{\hat n})\,  \kI ,
\nonumber \\
\vs^1\mp i \vs^2 &=&
\sum_{djm} \om^{d-4} \, \syjm{\pm4}{jm}(\mbf{\hat n})\,
\big(\kE\pm i\kB\big) ,
\nonumber \\
\vs^3 &=&
\sum_{djm} \om^{d-4} \, \syjm{}{jm}(\mbf{\hat n})\, \kV ,
\label{vac_exp}
\eea
where $|s|\leq j\leq d-2$.
The CPT-odd operators in the Lagrange density
are controlled by the coefficients $\kV$.
The dimension $d\geq 4$ is even
for the spherical coefficients $\kI$ for Lorentz violation,
while $d\geq 5$ is odd for $\kV$
and $d\geq 6$ is even for $\kE$ and $\kB$.
The number of independent components for each of $\kI$ and $\kV$
is $(d-1)^2$,
and the number for each of $\kE$ and $\kB$
is $(d-1)^2-16$.
In the language of spherical coefficients,
anisotropic effects are governed by all coefficients with $j\neq 0$,
dispersion manifests for all coefficients except $\kIdjm 4{jm}$,
while birefringence occurs for all coefficients except $\kI$.
We remark in passing that this implies birefringence for even $d$
can occur only for nonminimal operators, $d\geq 6$,
unlike the case of Lorentz violation in the pure-photon sector
for which minimal $d=4$ birefringent operators exist
\cite{ck}.

Armed with the above tools,
we can use the gravitational-wave event GR150914
to test local Lorentz invariance in gravity.
Since operators with larger $d$ are expected to be more suppressed,
it might seem natural to study first 
effects involving the coefficients $\kIdjm 4{jm}$.
However,
the corresponding operators are nondispersive and nonbirefringent,
so detecting their effects is more challenging
and typically would require a comparison
to light or neutrinos propagating from the same source
\cite{kt15,ni16}. 
Instead,
we consider in turn the $d=5$ coefficients $\kVdjm 5{jm}$
and the $d=6$ cofficients $\kEdjm 6{jm}$, $\kBdjm 6{jm}$.
Since polarimetric information for GW150914 and its source is unavailable,
we focus here on dispersive effects.

Consider first the generic situation 
involving a source producing gravitational waves,
such as the merger and ringdown of a black-hole binary.
We can reasonably assume the observed wave 
is generated in a superposition of the two propagating modes.
For example,
the eigenmodes in the presence of $d=5$ Lorentz violation 
are circularly polarized,
so special physical circumstances would be required
for a source to produce only one eigenmode. 
The dispersion between the two modes
evinced in the dispersion relation \rf{disprel}
can then be used to constrain Lorentz violation
by comparing their arrival times.
The difference in their velocities 
generically depends on both the frequency $\om$ of the wave
and the location $\mbf{\hat n}$ of its source.
Both the source and the detector can be taken as comoving objects. 
The coordinate interval between them is therefore
$dl_c = (1+z) dl_p = -v_z dz / H_z$,
where 
$v_z$ is the velocity of the source at redshift $z$,
and
$H_z = H_0(\Om_r\ze^4 + \Om_m\ze^3
+\Om_k\ze^2+\Om_\La)^{1/2}$,
$\ze \equiv 1+z$,
is the Hubble expansion rate at $z$ expressed in terms of the
Hubble constant $H_0 \simeq67.3$ km/s/Mpc
radiation density $\Om_r\simeq 0$,
matter density $\Om_m\simeq 0.315$,
vacuum density $\Om_\La\simeq 0.685$, 
and curvature density 
$\Om_k = 1-\Om_r-\Om_m-\Om_\La$.
Although the coordinate distance is identical for both modes,
their velocities and hence their travel times 
differ in the presence of Lorentz violation.
For example,
the arrival-time difference between the two modes
for coefficients $\kV$ with fixed $d$ 
is given by
\beq
\De t \approx 2 \om^{d-4} 
\int_0^z \fr{(1+z)^{d-4}}{H_z}dz
\sum_{jm}
\syjm{}{jm}(\mbf{\hat n}) \kVdjm d {jm}.
\label{det}
\eeq

For GW150914,
the linear combinations of 
$\kVdjm 5{jm}$ or $\kEdjm 6{jm}$, $\kBdjm 6{jm}$
appearing in the corresponding expression for $\De t$
are determined by spin-weighted spherical harmonics
with approximate angular arguments 
$\th\simeq 160^\circ$, $\ph \simeq 120^\circ$
in the Sun-centered frame.
The source is located at redshift $z=0.09^{+0.03}_{-0.04}$
\cite{ligo}.
At the maximum amplitude of the observed chirp signal,
the width of the peak is approximately 0.003 s
and no indication of splitting is evident.
We can therefore reasonably take $\De t \lsim 0.003$ s.
The frequency $f=\om/2\pi$ of the chirp spans the range 35-250 Hz
\cite{ligo},
and we can adopt a conservative value of $f\simeq 100$ Hz.

With the above values,
we obtain the constraints 
\bea
&&
\hskip -30pt
\Big | 
\sum_{jm} \syjm{}{jm}(160^\circ, 120^\circ) \kVdjm 5 {jm} 
\Big | 
\lsim 
2 \times 10^{-14} {\rm ~m},
\label{kvconst}
\\
&&
\hskip -30pt
\Big | 
\sum_{jm} \syjm{\pm 4}{jm}(160^\circ, 120^\circ) 
\big(\kEdjm 6 {jm} \pm i\kBdjm 6 {jm} \big) 
\Big | 
\nonumber\\
&&
\hskip 120pt
\lsim 
8 \times 10^{-9} {\rm ~m}.
\label{kebconst}
\eea
The result \rf{kvconst} represents the first constraint
on pure-gravity Lorentz-violating operators with $d=5$.
It also represents the first limit 
on CPT violation in gravitational waves,
which here corresponds to a difference in propagation speed
between the two circularly polarized CPT-conjugate eigenmodes. 
Note that sensitivity to $d=5$ effects is empirically unavailable 
in the nonrelativistic limit 
and hence to typical laboratory experiments
on Newton gravity,
because the presence of $d=5$ Lorentz-violating operators 
in the action leaves unaffected Newton's law
\cite{bkx}.
This fact underscores the added value 
of the discovery of gravitational waves
in the context of studies of the foundations of relativistic gravity.
The result \rf{kebconst} represents the first bound
on all birefringent coefficients at $d=6$
and is competitive with existing laboratory bounds
\cite{lo15, hust15,hustiu}.

\begin{table}
\begin{center}
\tabcolsep8pt
\begin{tabular}{cc||c|c}
$d$ & $j$ & Coefficient & Constraint\\
\hline
\hline																																			
5	&	0	&	$	|	\kVdjm{	5	}{	0	0	}	|	$												&	$	<	6	\times	10^{	-14	}	$	m	\\
\hline																																			
5	&	1	&	$	|	\kVdjm{	5	}{	1	0	}	|	$												&	$	<	4	\times	10^{	-14	}	$	m	\\
	&		&	$	|	\kVdjm{	5	}{	1	1	}	|	$												&	$	<	1	\times	10^{	-13	}	$	m	\\
\hline																																			
5	&	2	&	$	|	\kVdjm{	5	}{	2	0	}	|	$												&	$	<	3	\times	10^{	-14	}	$	m	\\
	&		&	$	|	\kVdjm{	5	}{	2	1	}	|	$												&	$	<	7	\times	10^{	-14	}	$	m	\\
	&		&	$	|	\kVdjm{	5	}{	2	2	}	|	$												&	$	<	4	\times	10^{	-13	}	$	m	\\
\hline																																			
5	&	3	&	$	|	\kVdjm{	5	}{	3	0	}	|	$												&	$	<	3	\times	10^{	-14	}	$	m	\\
	&		&	$	|	\kVdjm{	5	}{	3	1	}	|	$												&	$	<	4	\times	10^{	-14	}	$	m	\\
	&		&	$	|	\kVdjm{	5	}{	3	2	}	|	$												&	$	<	2	\times	10^{	-13	}	$	m	\\
	&		&	$	|	\kVdjm{	5	}{	3	3	}	|	$												&	$	<	1	\times	10^{	-12	}	$	m	\\
\hline																																			
6	&	4	&	$	|	\kEdjm{	6	}{	4	0	}	|	$	,	$	|	\kBdjm{	6	}{	4	0	}	|	$	&	$	<	1	\times	10^{	-6	}	$	m$^{2}$	\\
	&		&	$	|	\kEdjm{	6	}{	4	1	}	|	$	,	$	|	\kBdjm{	6	}{	4	1	}	|	$	&	$	<	3	\times	10^{	-7	}	$	m$^{2}$	\\
	&		&	$	|	\kEdjm{	6	}{	4	2	}	|	$	,	$	|	\kBdjm{	6	}{	4	2	}	|	$	&	$	<	6	\times	10^{	-8	}	$	m$^{2}$	\\
	&		&	$	|	\kEdjm{	6	}{	4	3	}	|	$	,	$	|	\kBdjm{	6	}{	4	3	}	|	$	&	$	<	2	\times	10^{	-8	}	$	m$^{2}$	\\
	&		&	$	|	\kEdjm{	6	}{	4	4	}	|	$	,	$	|	\kBdjm{	6	}{	4	4	}	|	$	&	$	<	1	\times	10^{	-8	}	$	m$^{2}$	\\
\end{tabular}
\vskip -5pt
\caption{\label{table2}
Constraints on coefficients for Lorentz violation.}
\end{center}
\vskip -10pt
\end{table}

Some insight into the implications of these bounds
can be gained by deriving from them 
the constraint on each individual coefficient in turn,
under the assumption that the other components vanish. 
The resulting estimated bounds 
on the modulus of each component of 
$\kVdjm 5 {jm}$, $\kEdjm 6{jm}$, and $\kBdjm 6{jm}$
are displayed in Table \ref{table2}. 
Note that each entry thereby also represents constraints 
on the moduli of the real and imaginary parts of each component.
This standard practice 
\cite{tables}
is useful
in comparing limits across different experiments
and 
in constraining specific models.
For example,
models with rotation-invariant gravitational Lorentz violation 
\cite{mi12,bl15}
can involve at most the spherical coefficients 
$\kI$ and $\kV$ with $jm = 00$,
for which it is convenient to define
$\kIring \equiv \kIdjm{d}{00}/{\sqrt{4\pi}}$
and $\kVring \equiv \kVdjm{d}{00}/{\sqrt{4\pi}}$.
The rotation-invariant limit of the dispersion relation \rf{disprel}
then takes the form
\beq
\om = 
\big (1 - \kIdring 4 \big) |\pvec| 
\pm \kVdring 5 \om^2 - \kIdring 6 \om^3
\pm \kVdring 7 \om^4 - \kIdring 8 \om^5
\pm \ldots ,  
\eeq
and the first row of Table \ref{table2} constrains $\kVdring 5$.
Note that the $\pm$ signs reflect the presence of birefringence
and CPT violation for even powers of $\om$,
both of which are required to describe physics 
associated with an effective field theory
\cite{km13}.

In principle,
methods related to the one adopted here 
could be used to obtain estimated constraints 
on other coefficients with $d>5$,
which are all associated with dispersive operators. 
The approach used above can be applied
directly to $\kE$, $\kB$, and $\kV$,
as these always control birefringent operators.
For example,
it yields the approximate bounds
$|\kVdjm 7 {jm}| \lsim 1 \times 10^{-2}$ m$^{3}$.
In contrast,
no birefringence occurs for $\kI$,
so a dispersive analysis for this type of Lorentz violation
requires a somewhat different approach.
One option might be to reverse-propagate the observed signal to the source 
while allowing for the presence of frequency-dependent Lorentz violation,
comparing the result 
to waveform templates for black-hole coalescence to extract constraints.
The resulting limits on the coefficients $\kI$ for $d=6,8$
would be significantly weaker 
than ones already deduced from the absence 
of gravitational \cv\ radiation in cosmic rays
\cite{kt15}.
Note,
however,
that dispersion limits of the type discussed here are particularly clean 
because they involve comparing the properties of two gravitational modes
and hence lie entirely within the pure-gravity sector,
whereas bounds from gravitational \cv\ radiation
involve comparative tests between the gravity and matter sectors.
Indeed,
gravitational \cv\ radiation may even be forbidden
for certain relative sizes of the coefficients 
for Lorentz violation for gravity and matter,
which would obviate any bounds obtained via this technique.
We also note in passing that the results in Ref.\ \cite{kt15}
are presented as limits on components of $\cHat{}^{\mn}{}_{\mn}$,
but the analysis effectively bounds $\vs^0$ and hence $\kI$,
which contains pieces of both 
$\cHat{}^{\mn}{}_{\mn}$ and $\dHat{}^{\mn}{}_{\mn}$.

The future detection of additional gravitational-wave events
will yield direct improvements in the constraints obtained in this work.
Moreover,
the use of dispersion information from multiple astrophysical sources 
at different sky locations
permits extraction of independent constraints on different coefficients,
as has already been demonstrated 
for noniminal coefficients in the photon and neutrino sectors of the SME
\cite{photdisp,nudisp}.
Improved sensitivities can also be expected
for gravitational waves of higher frequency,
as might be emitted in a supernova core collapse.
The prospects are evidently bright 
for future studies of foundational physical principles 
via measurements of gravitational-wave properties.

\medskip

This work was supported in part 
by the United States Department of Energy
under grant number {DE}-SC0010120,
by the United States National Science Foundation 
under grant number PHY-1520570,
and by the Indiana University Center for Spacetime Symmetries.

\end{document}